\documentclass[twocolumn,floats,showpacs,superscriptaddress,pre]{revtex4}

\expandafter\let\csname equation*\endcsname\relax

\expandafter\let\csname endequation*\endcsname\relax

\usepackage{amsmath}

\usepackage{graphicx,psfrag,bbm,latexsym,color,dcolumn,bm,dsfont,bbm,
color,mathrsfs,bbold,latexsym,amsfonts,amssymb}

\usepackage{hyperref} 

\usepackage{amssymb}

\definecolor{myred}{RGB}{168,5,14}
\definecolor{myblue}{RGB}{13,13,255}
\definecolor{mygreen}{RGB}{20,150,20}

\begin{document}

\title
 { {Absence of small-world effects at the quantum level and\\
stability of the quantum critical point}}

\author{Massimo Ostilli}
\affiliation{Instituto de F\'isica,  Universidade Federal da Bahia, Salvador 40170-115, Brazil}


\begin{abstract}
  The small-world effect is a universal feature used to explain many different
  phenomena like percolation, diffusion, and consensus.
  Starting from any regular lattice of $N$ sites,
  the small-world effect can be attained by rewiring randomly an $\mathcal{O}(N)$ number of links
  or by superimposing an equivalent number of new links onto the system.
  In a classical system this procedure is known to change radically its critical point and behavior, the new system
  being always effectively mean-field.
  Here, we prove that at the quantum level the above scenario does not apply:
  when an $\mathcal{O}(N)$ number of new couplings are randomly superimposed onto
  a quantum Ising chain, its quantum critical point and behavior both remain unchanged.
  In other words, at zero temperature quantum fluctuations destroy any small-world effect.
This exact result sheds new light on the significance of the quantum critical point as a thermodynamically stable feature
of nature that has no analogous at the classical level and essentially prevents a naive application
  of network theory to quantum systems.
The derivation is obtained by combining the quantum-classical mapping with a simple topological argument.
\end{abstract}

\maketitle

\email{massimo.ostilli@gmail.com}

\section{Introduction}
After Onsager' s exact solution for the D=2-dimensional case~\cite{Onsager},  
the Ising model assumed the role of a paradigm at the base of our understanding of phase transitions and critical
phenomena, not only within classical physics, but also within quantum physics.
However, important conceptual and quantitative differences exist between the two cases.
In the classical case, 
competition between order and thermal fluctuations
induces a second-order phase transition triggered by lowering the temperature
below a critical value which is finite if D$\geq$ 2.
In the quantum case, 
at zero temperature, the competition between two different ground states
induces a second-order phase transition triggered by lowering a transverse
external field below a critical value which is finite even when D=1~\cite{Sondhi,Sachdev,Vojta,Essler,Dutta}.
Indeed, as the quantum-classical mapping (QCM) shows~\cite{QCM},
the critical behavior of a D-dimensional quantum Ising model amounts to that of a suitable D~$+1~$ classical Ising model.
Despite these well established facts and an extensive literature on the subject, many issues remain open about the interplay
between the classical and the quantum case~\cite{Sachdev}.

  \begin{figure}[h]
  \centering
  \includegraphics[width=0.8\columnwidth,clip]{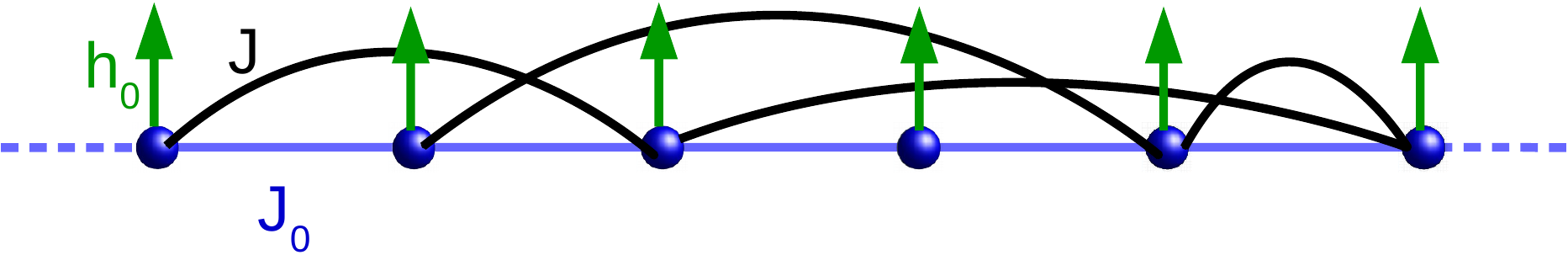}
  \caption{ 
    A portion of the model studied in this paper: 
    six qubits in a chain are immersed in a uniform transverse field $h_0$ (green arrows) and  
    interact by nearest neighbors 
    and random pairs via the ferromagnetic couplings $J_0$ (blue links) and $J$ (black links), respectively.
    The mean connectivity related to the coupling $J$ is denoted by $c$.
    {When $h_0\to 0$, we recover a classical Ising model whose underlying graph,
      despite being embedded in a one-dimensional space, is characterized by the SW effect and, as a consequence,
    makes the critical behavior of the classical model mean-field. However, does such a classical scenario hold also in the quantum case, i.e., when $h_0\neq 0$?}
  }
  \label{fig3}
\end{figure}

  \begin{figure}[h]
  \centering
  \includegraphics[width=0.5\columnwidth,clip]{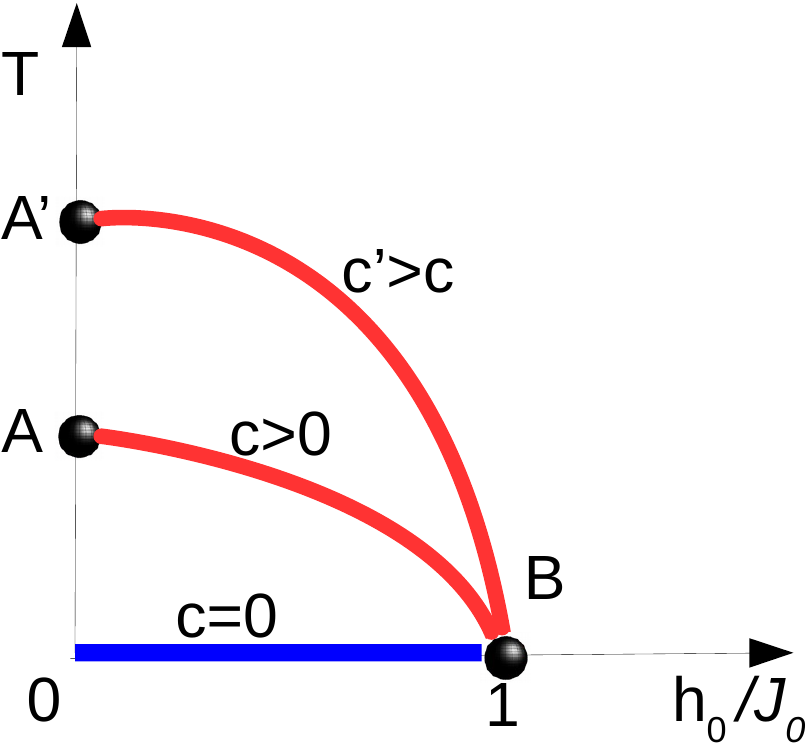}
  \caption
      {
    Phase diagram 
    of the quantum Ising chain of Fig. \ref{fig3},
    Eq. (\ref{H}), on the plane $(h_{0}/J_{0},T)$, where $J_0$ is the coupling and $h_0$ is the transverse field.
    The curves $c>0$ and $c'>c$ (red), represent qualitative lines of critical temperature for two different values of
    the additional mean connectivity (bringing the same long-range coupling $J$),
    $c$ and $c'$, as a function of the adimensional parameter $h_{0}/J_{0}$.
    For $c=0$ the line of critical temperature collapses to 0 (blue).
    The points $A$ (or $A'$) and $B$ represent
    classical and quantum critical points, respectively, and can be exactly calculated: $A\equiv(0,T_{c}(h_{0}=0))$,
    $T_{c}(h_{0}=0)$ being the solution of Eq. (\ref{h00}); while $B\equiv(1,0)$ for any $c$, being this the main result of the present work. 
    The critical behavior in $A$ (or $A'$) is mean-field while that in $B$ corresponds to that of a {D=2} Ising model.
    As we increase $c$ toward $c'$, the point $A$ moves upward toward $A'$, but $B$ does not move at all.
  }
  \label{fig1}
\end{figure}

Aside from the finite dimensional case, the classical Ising model and its generalizations have extensively been used
also in the context of networks (complex or not), where the nodes and the links of the network 
represent the spins and the couplings between them, respectively~
\cite{Parisi,Fisher,Watts,Bollobas,Zecchina,Barabasi,DM,NewBar,Guido,Review}.
In these systems, the phase transition is seen as a result of the interplay between the
topology of the underlying network, ranging from completely regular to completely random, and the physical process running on it.
Whenever the topology is sufficiently random these complex systems share a common feature: their dimension is effectively infinite
and, as a consequence, the system is effectively mean-field.
This property can be understood by exploiting the small-world (SW) concept~\cite{Watts}.
  In fact, the general feature of random graphs and their variants, called SW graphs, is their SW character:
  as opposed to D-dimensional lattices, where
  the average distance between two randomly chosen nodes scales as $N^{1/\mathrm{D}}$, in SW graphs
  the distance scales as $\log(N)$, from which, by comparison we see that in the latter case D~$=\infty$.
  In terms of phase transitions, this SW property results 
  in dramatically favoring long range order, the system being effectively mean-field due to the infinite-dimensionality~\cite{Review}.
  However, the above scenario applies to classical systems and a
  natural question emerges: Does it hold for quantum systems too?
  Indeed, a bunch of speculative works trying to exploit the SW effect as well as the general concepts of complex networks
  at the quantum level have appeared in recent years~\cite{Biamonte}
  in which, e.g., Anderson localization is reduced~\cite{Anderson}, entanglement is improved~\cite{Entanglement},
  super conductivity is enhanced~\cite{BianconiSC},
  and even the way to a visionary quantum internet is paved~\cite{QuantumInternet}, to mention a few.
  Our finding, however, which is based on an exact result at zero temperature, leads 
  to skepticism.

  Let us consider specifically the SW graphs:
  Starting from any regular D-dimensional lattice, the SW property can be attained by rewiring
  an $\mathcal{O}(N)$ number of couplings or by superimposing an equivalent number of extra couplings onto the original system.
  The latter procedure generates more analytically treatable models and several exact results have been reached
 ~\cite{Hastings,Nikoletopoulos}, also for the $\mathrm{D}\geq 2$ case~\cite{SW,SW2}.
  In particular, for the classical Ising model these studies show that, due to the SW effect,
  in the {D=1} case the system acquires a finite critical temperature, while in the {D=2} case
  the system gets a higher critical temperatures
  and, in both cases, the critical behavior turns out to be mean-field like.

  Here we prove that this classical scenario does not hold at the quantum level:
  when an $\mathcal{O}(N)$ number of extra ferromagnetic couplings are randomly superimposed onto
  a {D=1} quantum Ising chain, see Fig. \ref{fig3}, its quantum critical point and behavior both remain unchanged, see Fig. \ref{fig1}.
  In other words, at zero temperature quantum fluctuations destroy any SW effect.
  As we shall show, the ultimate reason for that is the fact that, at zero temperature,
  an extra dimension - the one associated with the imaginary time evolution -
  arises that is not covered by the SW links: the ``quantum graph'' is never SW.
  As a consequence, caution is in order before transferring the established knowledge of classical complex systems into the
  quantum world.
  This exact result sheds new light on the meaning of the quantum critical point as a thermodynamically stable feature 
  of systems and, as we explain later, provides insights for understanding the intrigued interplay between quantum and classical behavior
  at finite temperatures in the presence of a transverse field~\cite{Sachdev}.
  Furthermore, the stability of the quantum critical point 
  might be crucial for determining optimal quantum annealing paths ~\cite{Nishimori,Farhi,Farhi2,Santoro,Morita,Das,Lidar}
  for hard combinatorial problems~\cite{Battaglia,Krzakala,Krzakala2,Hen,Denchev,Mandra,Mandra2,Mandra3,Katzgraber,QREM}.
  The main aim of the present work however is to focus on the proof of the result. {The first part of this result, i.e., the fact that
    there is no SW effect in a quantum system, will be first made evident by using a quite simple geometrical argument based on the quantum-classical mapping
    (see top and middle panels of Fig. \ref{fig2}). Then, the second part of the result, i.e., the fact that the critical point remains unchanged
    when the random extra couplings are added, will be derived by using another special mapping that sends
    a random model toward a non random one (see bottom panel of Fig. \ref{fig2}). Furthermore, it will be shown that this second mapping confirms also that the critical behavior of
    the system remains unchanged: a fact that represents a necessary and sufficient condition for the absence of any SW effect and which is in agreement
    with the above simple geometrical argument.}    

  The paper is organized as follows:
  In Sec. II we recall the QCM and its immediate application to the one-dimensional model without disorder (the pure model).
  In Sec. III we discuss the crucial difference of the QCM at zero and finite temperature (not often stressed in the literature).
  In Sec. IV we introduce the random model to which we apply the QCM followed by the application
  of the random-non-random mapping (RNRM), which eventually produces a model without disorder.
  In Sec. V we solve this model at $T=0$. In Sec. VI we discuss the state of the art at $T>0$. Finally,
  in Sec. VII conclusions are drawn.
  The paper is equipped with an Appendix where we report and revisit the derivation of the RNRM, originally given in Ref. \cite{MOI}, 
  and provide also an alternative derivation.
  
  \section{The pure Model and The Quantum-Classical Mapping (QCM)}
  Let us consider a lattice ring of $N$ qubits
  interacting via a first neighbor ferromagnetic coupling $J_0>0$ and subjected to a transverse field $h_0$
\begin{eqnarray}
  \label{H0}
  H_0=-J_0\sum_{i=1}^N\sigma_i^Z\sigma_{i+1}^Z-h_0\sum_{i=1}^N\sigma_i^X,
\end{eqnarray}
where $\sigma_i^X$, $\sigma_i^Y$ and $\sigma_i^Z$ are the Pauli matrices of the $i$-th qubit.
This system is known to develop a zero temperature second-order quantum phase transition
at the critical point $J_0=h_0$. A way to see this consists in solving the model via the Jordan-Wigner transformations
~\cite{Pfeuty,SuzukiBook}.
Another interesting way consists in applying the QCM~\cite{QCM}.
For $\beta\to\infty$, the QCM evaluates the partition function
$Z_0=\mathrm{Tr}\exp(-\beta H_0)$ of the quantum {D=1} model with Hamiltonian (\ref{H0}),
as the partition function of an anisotropic classical {D=2} Ising model defined by
two suitable couplings associated with two directions $x$ and $y$ (not to be confused with the suffix of the Pauli matrices):
$x$ corresponds to the position of the actual $i$-th qubit, and $y$
corresponds to a virtual direction along which we propagate the inverse temperature $\beta$ (or, equivalently, the
imaginary time). The resulting classical Hamiltonian reads (for a review see e.g. Refs.~\cite{Sachdev,Dutta})
\begin{align}
  \label{H0cl}
  H_0^{Classic}=&-J_{0x}\sum_{j=1}^M\sum_{i=1}^NS_{i,j}S_{i+1,j}\nonumber \\
  & -J_{0y}\sum_{i=1}^N\sum_{j=1}^MS_{i,j}S_{i,j+1},
\end{align}
where: the $S_{i,j}$ are $MN$ virtual classical spins arranged onto the
{D=2} discrete torus $[1,\ldots,M]\times [1,\ldots,N]$ 
and, up to terms $\mathcal{O}(1/M^2)$, the two couplings are given as follows
\begin{eqnarray}
  \label{JJ}
  &J_{0x}=\frac{J_0}{M}, \quad
  \beta_0 J_{0y}=\frac{1}{2}\ln\left(\frac{M}{\beta_0 h_0}\right),
\end{eqnarray}
where $\beta_0=1[\beta]$ is a unitary constant that serves only for dimensional reasons ($[\beta_0]=[\beta]$). 
Systems (\ref{H0}) and (\ref{H0cl}) become equivalent in the limit $M\to\infty$, i.e, in the limit in which the Trotter-Suzuki
factorization~\cite{QCM},
at the base of the QCM, becomes exact. In this limit, $J_{0x}\to 0^+$ while $J_{0y}\to+\infty$ in such a way that the system can have a finite
critical point. In fact, by plugging Eqs. (\ref{JJ}) into the equation for the critical point of the {D=2} Ising model
(from Kramers-Wannier duality~\cite{Kramers}, or Onsager' s solution~\cite{Onsager}), 
\begin{eqnarray}
  \label{Onsager}
  \sinh(2\beta_0 J_{0x})\sinh(2\beta_0 J_{0y})=1,
\end{eqnarray}
in the limit $M\to\infty$, we get the critical point of the original {D=1}
quantum system (\ref{H0}): $J_0=h_0$.

\section{QCM at zero and finite temperature - Technical warnings}
In the following sections, 
the reader should take into account two technical warnings not often stressed in the literature. 
These warnings are related to the crucial difference that exists between the application of QCM at zero and finite temperature:

(i) As first remarked by Suzuki~\cite{QCM}, 
the QCM does not claim that the critical behavior of the D-dimensional quantum system $H_0$ at finite temperatures
is equal to that of the D~$+$~1-dimensional classical system $H_0^{Classic}$ since, for any finite $\beta$, the latter, due to Eqs. (\ref{JJ}),
degenerates when $M\to\infty$ so that, in general, at finite temperature, the resulting classical model might be equivalent
to a suitable classical - but rather non obvious - D-dimensional model.
Let us analyze this issue more closely. It is easy to see that, in the thermodynamic limit, the ground
state energy $E_0$ of the system (\ref{H0}) can be expressed as
\begin{align}
  \label{E0}
  &\lim_{N\to\infty}\frac{E_0}{N}=-\lim_{N\to\infty}\lim_{\beta\to\infty}\frac{1}{N\beta}\log\left[\mathrm{Tr}\exp(-\beta H_0)\right]
  \nonumber \\
  &=-\frac{1}{\beta_0}\lim_{M,N\to\infty}\frac{1}{MN}\log\left[\mathrm{Tr}\exp(-\beta_0 H_0^{Classic})\right],
\end{align}
where the last equality holds up to an immaterial additive constant that does not depend on the Hamiltonian parameters
(see Sec. IV of Ref.~\cite{QCM}).
From Eq. \ref{E0} it is then evident that, apart from the physical dimension, the length $M$ plays the same role of the inverse
temperature in the limit in which this goes to infinity and, as a consequence, the zero temperature limit is also the limit $M\to\infty$
where
the Trotter-Suzuki factorization becomes exact. At finite temperature, one can still exactly exploit the Trotter-Suzuki
factorization, but the resulting classical system system does not acquire an actual extra dimension.
In fact, at finite temperature
one has to evaluate the free energy density $f_0$ as
\begin{align}
  \label{f0}
  -\beta f_0 & =\lim_{N\to\infty}\frac{1}{N}\log\left[\mathrm{Tr}\exp(-\beta H_0)\right]\nonumber \\ &=
  \lim_{N\to\infty}\frac{1}{N}\log\left[\lim_{M\to\infty}Z_0^{Classic}\right],
\end{align}
which is very different from $\lim_{M,N\to\infty}\frac{1}{MN}\log\left[Z_0^{Classic}\right]$.
The finite temperature analysis is therefore rather harder because
it requires the exact solution of the D=2 model for finite sizes where one of the two sides of the torus must be sent to
infinity while keeping the other fixed.
In particular, as shown in Ref.~\cite{QCM},
for the model (\ref{H0}) it is possible to obtain $f_0$ by exploiting the exact solution of the D=2 Ising model for finite sizes from
Kaufman~\cite{Kaufman} and the result is
\begin{eqnarray}
  \label{finiteT}
  -\beta f_0=\frac{1}{2\pi}\int_0^{2\pi}dq\log\left[2\cosh\left(\beta \epsilon(q)\right)\right],
\end{eqnarray}
where
\begin{eqnarray}
  \label{finiteT1}
  \epsilon(q)=(J_0^2+h_0^2-2J_0h_0\cos(q))^{1/2}.
\end{eqnarray}
Remarkably, Eqs. (\ref{finiteT})-(\ref{finiteT1}) cannot be obtained from the thermodynamic limit of the free energy density of the
D=2 classical Ising
model by simply using in it the couplings (\ref{JJ}) in the limit $M\to\infty$. Notice in particular that, as anticipated, 
for any finite $\beta$ Eqs. (\ref{finiteT})-(\ref{finiteT1}) do not show any singularity in the Hamiltonian parameters or in $\beta$.
Only for $\beta\to\infty$ does a singularity show up at the critical point $J_0=h_0$ and only in this limit is one allowed to exploit 
the thermodynamic limit of the free energy density of the D=2 classical Ising
model by simply using in it the couplings (\ref{JJ}) in the limit $M\to\infty$.
Later on, we shall need to work with the free energy density of the D=2 classical Ising
model but in the presence of a row dependent external field.

ii) Provided the result is used only for zero temperature (see warning (i)), the QCM makes sense and is useful for analytic manipulations also for large but finite $M$,
provided that $M$ scales at least proportionally with $N$. As a practical rule one can simply take $M=N$.

We stress again that in the following analysis
we are only concerned with the zero temperature limit and the symbol $\beta_0$ plays no role but
  a mere dimensional factor.

\begin{figure}[t]
  \centering
  \includegraphics[width=0.9\columnwidth,clip]{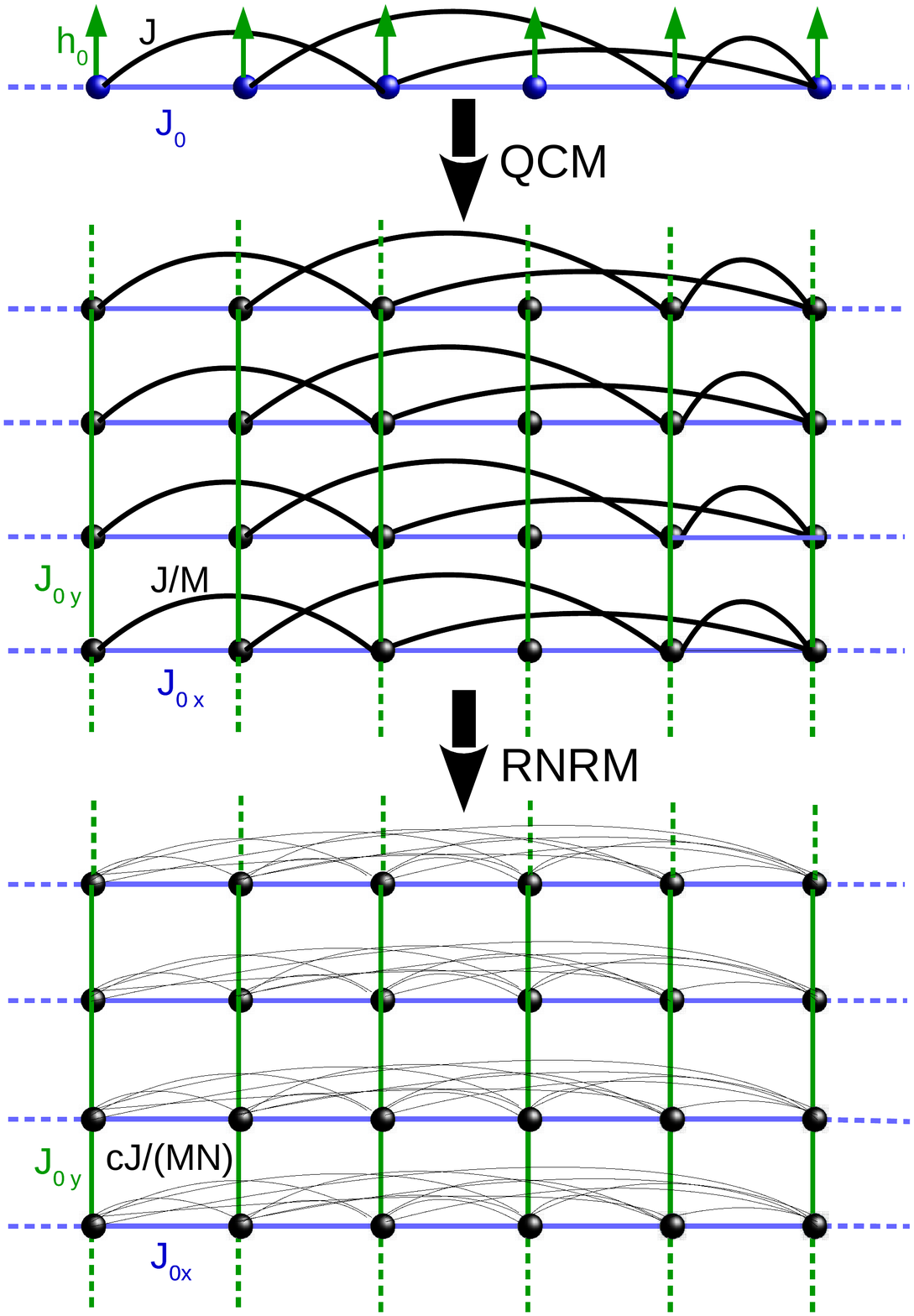}
  \caption
      {
    Application of the quantum-classical mapping (QCM) followed
    by the random-non-random mapping (RNRM) to the quantum Ising chain of Eq. (\ref{H}). Here $M=4$ and $N=6$.
    Dashed lines represent
    periodic boundary conditions. 
    Top: 
    Six qubits in a chain are immersed in a uniform transverse field $h_0$ (green arrows) and  
    interact by nearest neighbors 
    and random pairs via the ferromagnetic couplings $J_0$ (blue links) and $J$ (black links), respectively.
    {Notice that, in the classical limit, i.e., in the case with a null transverse field, $h_0=0$, the underlying network, even if embedded in a 1D space, makes the system SW.}   
    Middle: Array
    of $4\times 6$ classical spins interacting by nearest neighbors
    via the couplings $J_{0x}$ (blue links) and $J_{0y}$ (green links), as given by Eqs. (\ref{JJ}),
    and by random pairs via the coupling $J/M$ (black links). {Notice the absence of long-range links in the vertical direction: a fact that
      prevents the underlying network, now embedded in a 2D space, to make the system SW.}
    Bottom: Array
    of $4\times 6$  classical spins each interacting with nearest neighbors
    by the couplings $J_{0x}$ and $J_{0y}$, and with all the other spins lying on the same row
    by the uniform coupling $c J/(MN)$, as given by Eq. (\ref{RNR}) for large $M$.
    {This classical model can be exactly solved by using the exact solution of the 2D Ising model in the presence of a uniform external field (see Sec. V).}
  }
    \label{fig2}
\end{figure}

  \section{The Model with Small-World couplings and The Random-Non-Random Mapping (RNRM)}
Let us now see what happens when we add $cN$ extra interactions
between random pairs of qubits via another ferromagnetic coupling $J>0$; with $c>0$ being the additional mean connectivity;
see the upper panel of Fig. \ref{fig2}.
The new quantum Hamiltonian reads
\begin{align}
  \label{H}
  H=& -J\sum_{i< j=1}^Nc_{i,j}\sigma_{i}^Z\sigma_{j}^Z
  -J_0\sum_{i=1}^N\sigma_i^Z\sigma_{i+1}^Z
  -h_0\sum_{i=1}^N\sigma_i^X,
\end{align}
where $c_{i,j}$ is the adjacency matrix of the extra couplings, i.e., a random variable taking the values 0 or 1 with
probabilities $1-c/N$ or $c/N$, respectively. The resulting ensemble of graphs generated by the different realizations of $\{c_{i,j}\}$
is known as the Gilbert random graph (a slight variant of the Erd\"os-Reny random graph),
and its properties are well known~\cite{Bollobas}. In particular, for large $N$, the connectivity of each node
becomes Poissonian distributed with mean $c$ and, for any $c>1$,
the graph is percolating and owns the SW property. In general,
for different realizations of $\{c_{i,j}\}$ there correspond different Hamiltonians (\ref{H}).
Yet, in the thermodynamic limit $N\to\infty$, due to the self-averaging character
of the random graph, relative fluctuations of the system become negligible.
In other words,
any extensive observable (like the energy or the magnetization),
can be evaluated either via a single realization of a sufficiently large system
or as an average over the adjacency matrix realizations. 
The latter is the usual successful set-up
applied to all (quenched~\cite{Parisi}) disordered models in classical physics which, thanks to the QCM,
we can assume to be valid also in the present quantum case.

On applying the QCM to the quantum system (\ref{H}) we get
\begin{align}
  \label{Hcl}
  H^{Classic}=&-\frac{J}{M}\sum_{j=1}^M\sum_{i_1< i_2=1}^N c_{i_1,i_2}S_{i_1,j}S_{i_2,j} + H_0^{Classic},
\end{align}
where $H_0^{Classic}$ is defined in Eqs. (\ref{H0cl})-(\ref{JJ}).
From the first term of Eq. (\ref{Hcl}), we see that the SW effect on the underlying graph of $H^{Classic}$, if any,
can be realized only along the x-direction; see the middle panel of Fig. \ref{fig2}. In other words, 
in order to cross the {D=2} torus $[1,\ldots,M]\times [1,\ldots,N]$ from one corner to the opposite {via a sequence} of random hoppings
on the underlying graph of $H^{Classic}$, on average
we must use a $\mathcal{O}(\log(N)M)$ number of links, while in a SW graph this should be $\mathcal{O}(\log(MN))$.
As a consequence, we expect that $H^{Classic}$, and hence the quantum system governed by $H$, has not acquired a mean-field character,
and that it remains essentially similar to the original quantum system governed by $H_0$ {with no SW effect}.
In the following we prove that this guess is exact in an extreme sense: not only the quantum critical behaviors,
but also the quantum critical points of $H_0$ and $H$ are the same.

Let us indicate the averages over the $\{c_{i_1,i_2}\}$ realizations by $\overline{\cdot}$.
As explained above, if $\langle \mathcal{O}\rangle=\mathrm{Tr}\mathcal{O}\exp (-\beta H)/\mathrm{Tr}\exp (-\beta H)$
denotes the ensemble average of the observable $\mathcal{O}$ associated with a given $\{c_{i_1,i_2}\}$ (quenched) realization,  
for $N$ large, we can conveniently identify this average with $\overline{\langle \mathcal{O}\rangle}$. In turn, all these averages
can essentially be derived from the free energy density $f$ of the quenched model:
$-\beta f=\lim_{N\to\infty} \overline{\ln(Z)}/N=\lim_{N\to\infty}\lim_{n\to 0}(\overline{Z^n}-1)/(Nn)$.
The latter identity is at the base of the so called replica-trick that has been used to investigate a large variety of random models,
especially spin-glass models~\cite{Parisi}. At the critical point, when the replica-trick 
is used in combination with the high temperature expansion of the free energy of a model built over a random graph,
there emerges a general mapping
between the random model (or ``disordered model'') and a suitable non random model~\cite{MOI,SW}.
In the following, we refer to this mapping as the random non random mapping (RNRM).
In general, the disorder can be due to the underlying graph structure having a generic random matrix $\{c_{i_1,i_2}\}$
and, more in general, to the random values of the corresponding couplings $\{J_{i_1,i_2}\}$. In both cases,
the RNRM consists of the following replacement (here $\overline{\cdot}$ means average over any kind of
disorder)
\begin{eqnarray}
  \label{RNR}
  c_{i_1,i_2} \tanh(\beta J_{i_1,i_2}) \to \overline{c_{i_1,i_2} \tanh(\beta J_{i_1,i_2})}. 
\end{eqnarray}
As has been confirmed also via Monte Carlo simulations~\cite{SW2},
whereas 
the RNRM (\ref{RNR}) gives only effective approximations below
the critical temperature, it provides the exact critical point and behavior by
ruling out all the difficulties of the random model. 
We stress that the RNRM does not consist of some annealed approximation.
One of the most interesting advantages of the RNRM lies in the fact it holds true for generalized infinite-dimensional graphs without
the necessity for such graphs to be locally tree-like or small-world, just as in the present case. In Appendix A
we report a simple derivation that is valid for the present case and show also an alternative derivation which does not invoke
the replica trick at all.
When applied to the adjacency matrix $\{c_{i_1,i_2}\}$ of the random graph with the constant coupling $J/M$, for $M$ large but finite,
Eq. (\ref{RNR}) sends the random classical Hamiltonian (\ref{Hcl}) to the following non random Hamiltonian
(see the bottom panel of Fig. \ref{fig2})
\begin{align}
  \label{Hcl1}
    \tilde{H}^{Classic}=&-\frac{c J}{MN}\sum_{j=1}^M\sum_{i_1< i_2=1}^N S_{i_1,j}S_{i_2,j}
  + H_0^{Classic}.
\end{align}

\section{Solution of the model at $T=0$}
The Hamiltonian $\tilde{H}^{Classic}$ represents a {D=2} Ising model with superimposed fully connected interactions
that run only within the rows of the torus $[1,\ldots,M]\times [1,\ldots,N]$.
Its partition function $\tilde{Z}^{Classic}$ of the model (\ref{Hcl1}) can be analyzed
by a standard technique~\cite{MF}. 
By introducing $M$ auxiliary independent Gaussian fields $\{x_l\}$, up to a $\mathcal{O}(1)$ term in the exponent, and up to an immaterial constant
of proportionality, we get
\begin{eqnarray}
  \label{Ztildecl}
\tilde{Z}^{Classic}\propto \int \prod_{j=1}^M dx_j e^{-N\beta_0 f(\{x_j\})},
\end{eqnarray}
\begin{align}
  \label{ftildecl}
  \beta_0 f(\{x_j\})=& \frac{c\beta_0 J}{M} \sum_{j=1}^M x_j^2 \nonumber \\ &
  +\beta_0 f_0\left(\beta_0 J_{0x},\beta_0 J_{0,y};\left\{\frac{c \beta_0 J}{M} x_j\right\}\right), 
\end{align}
where $f_0(\beta J_{0x},\beta J_{0,y};\{ \beta h_j\})$ is the free energy density of the {D=2} Ising model
with the couplings $J_{0x}$ and $J_{0y}$ in the presence of a row-dependent external field $\{h_j\}$.
Notice that, in order to avoid a pedant notation like $f_0(\beta J_{0x},\beta J_{0,y};\{ \beta h_j\};N,M)$ etc.,
in Eq. (\ref{Ztildecl}) and following, the harmless dependencies on finite size effects are left understood
but they should be kept in mind for the correct interpretation of the next equations. 
We stress that such dependencies on finite size effects are smooth and not particularly important since, unlike
the harder situation where one has to deal with finite temperatures (see warning (i)), here we have $N\sim M$.
The steepest descent method applied to Eq. (\ref{Ztildecl}) provides the following effective
mean-field equations for the row-dependent average magnetizations $m_j=\sum_{i=1}^N\langle S_{i,j}\rangle/N$ of Eq. (\ref{Hcl1})
\begin{align}
  \label{mtildecl}
  m_j =& m_{0;j}\left(\beta_0 J_{0x},\beta_0 J_{0,y};\left\{\frac{c \beta_0 J}{M} m_l\right\}\right),
  \nonumber \\ &
  \quad j=1,\ldots,M, 
\end{align}
where $m_{0;j}(\beta_0 J_{0x},\beta_0 J_{0,y};\{\beta_0 h_j\})$ is the magnetization along the $j$-th row of the {D=2} Ising model
with the couplings $J_{0x}$ and $J_{0y}$ in the presence of a row-dependent external field $\{h_j\}$.
Equations (\ref{mtildecl}) and following are valid up to $\mathcal{O}(\log(N)/N)$ corrections.
Let us focus on the thermodynamically dominant uniform solution $\{m_j=m\}$ (making $f(\{m_j\})$ a minimum). 
By deriving Eq. (\ref{mtildecl}) with respect to a uniform external field $\{\beta h_j=\beta h\}$,
we obtain the adimensional susceptibility of the system
\begin{eqnarray}
  \label{chitildecl}
  \chi =\frac{\chi_0\left(\beta_0 J_{0x},\beta_0 J_{0,y}; \frac{c \beta_0 J}{M} m \right)}
  {1-\frac{c \beta_0 J}{M}\chi_0\left(\beta_0 J_{0x},\beta_0 J_{0,y}; \frac{c \beta_0 J}{M} m \right)},
\end{eqnarray}
where $\chi_0(\beta J_{0x},\beta J_{0,y};\beta h)$ is the adimensional susceptibility of the {D=2} Ising model
with the couplings $J_{0x}$ and $J_{0y}$ in the presence of a uniform external field $h$.
The paramagnetic solution of Eq. (\ref{mtildecl}) is stable when the denominator of Eq. (\ref{chitildecl})
evaluated at $m=0$ is positive.
In other words, for finite $M$, the paramagnetic solution becomes unstable when
\begin{eqnarray}
  \label{chitildecl1}
  \frac{c\beta_0 J}{M}\chi_0(\beta_0 J_{0x},\beta_0 J_{0,y};0)=1.
\end{eqnarray}
Taking into account the critical point of the {D=2} model, Eq. (\ref{Onsager}),
Eq. (\ref{chitildecl1}) tells us that, for large but finite $M$, the critical point of the system (\ref{Hcl1}) is shifted
at higher temperatures or, in terms of the couplings $\beta_0 J_{0x}$ and $\beta_0 J_{0,y}$, is such that
$\sinh(2\beta_0 J_{0x})\sinh(2\beta_0 J_{0y})<1$.
Furthermore, for $M$ large but finite, Eq. (\ref{mtildecl})
tells us that the system is essentially mean-field like, with the classical critical exponents.
In fact, from Eq. (\ref{chitildecl}) we see that the susceptibility has the critical exponent $\gamma=1$ and
similarly for the other critical exponents.
However, the QCM holds true only in the limit $M\to\infty$.
In such a limit, Eq. (\ref{chitildecl1}) can be satisfied only
at the critical point of the {D=2} model (\ref{Onsager}) (where $\chi_0(\beta_0 J_{0x},\beta_0 J_{0,y};0)\to\infty$)
which, by using explicitly the expression for the couplings, Eqs. (\ref{JJ}),
implies that the critical point of the quantum system with extra random couplings, Eq. (\ref{H}),
is just equal to the critical point of the quantum system without random couplings, Eq. (\ref{H0}): $J_0=h_0$. Furthermore, if we choose
$J_0<h_0$, so that we are in the paramagnetic region, on sending $M\to\infty$ in Eq. (\ref{chitildecl}), we get
\begin{eqnarray}
  \label{chipara}
\lim_{M\to\infty }\chi =\lim_{M\to\infty }\chi_0\left(\beta_0 J_{0x},\beta_0 J_{0,y}; 0 \right),
\end{eqnarray}
which implies that the critical exponent of the susceptibility of the system (\ref{H})
is equal to the critical exponent of the susceptibility of the quantum system (\ref{H0}) ($\gamma=7/4$), and the same argument
applies to the other critical exponents.
Notice that our analysis does not necessarily imply that the systems (\ref{H0}) and (\ref{H}) at $T=0$ are identical.
As we have mentioned before, 
inside the ferromagnetic region, i.e., for $J_0>h_0$, the RNRM is only an approximation
where Eq. (\ref{mtildecl}) turns out to be effective~\cite{SW2}.

\section{State of the art at $T>0$}
The above result is limited to zero temperature.
The other available exact result concerns the pure classical model, i.e., the case with no transverse field, $h_0=0$.
As discussed in the introduction, in this case, for any $c>0$ the system is effectively mean-field and  
its critical point can be exactly calculated by the following equation
~\cite{Nikoletopoulos,SW} 
\begin{eqnarray}
  \label{h00}
  c\tanh(\beta J)e^{2\beta J_0}=1.
\end{eqnarray}
Equation (\ref{h00}) tells us that the critical temperature is a growing function of $c$ (linear for large $c$).
For any other case, i.e., the region $(h_0>0,T>0)$, there are no exact results, however,
as discussed before in warning (i), in this region the {D=1} quantum system (\ref{H0}) behaves essentially as
a {D=1} classical system (\ref{finiteT}) so that, analogously, we expect that for any $T>0$
the {D=1} quantum system with superimposed additional links (\ref{H}), behaves essentially as
a {D=1} classical system with superimposed additional links too,
having therefore the small-world character, a finite critical temperature, and mean-field critical exponents.
By using these observations and interpolating the exact points given by Eq. (\ref{h00}) and the quantum critical point,
we get the scenario depicted in Fig. \ref{fig1},
with a line of critical temperatures that grows with $c$ for any $h_0<J_0$.

For $T>0$ the critical behavior is classical for any value of the Hamiltonian parameters,
however, this does not imply that all the physics of the system
is dominated by a classical behavior. More precisely, sufficiently far from the critical line, things might be not classical even for $T>0$.
In fact,
when $c=0$, for $T>0$ two lines of a D=1-classical $\bm{\rightarrow}$ D=1-quantum crossover,
are known to exist that depart from the quantum critical point $h_0=J_0$
(for fixed $T$, the above crossing corresponds to transiting from $h_0<J_0$ to $h_0>j_0$)
~\cite{Sondhi,Sachdev,Vojta,Dutta}.
From the experimental point of view, this crossover region
represents the most important part of the problem~\cite{Sachdev}.
Our analysis leads one to expect that, for $c>0$, such a crossover might become mean-field-classical
$\bm{\rightarrow}$ D=1-quantum.
This is a rather interesting issue
that deserves further investigation. 

\section{Conclusions and Perspectives}
In conclusion, we have proved that, contrary to a somehow classical common sense,
at zero temperature there is no
SW effect, the quantum critical point and behavior of the system remaining those of the finite dimensional
model before the addition of the extra links or, equivalently, before the rewiring.
Quantum fluctuations destroy any SW effect
and raise the quantum critical point as a robust feature of nature.  
Whereas this invariance puts severe limits on the possibility for improving long range order via the SW effect at low temperatures, 
we expect possible applications for quantum annealing.

It is natural to ask whether the stability of the quantum critical point remains valid also for more general
  networks, in particular those that are scale free. This will be the subject of a future work. It is should be, however, clear
  from our discussion based on the QCM (the same argument applies unchanged),
  that the small-world property cannot be satisfied regardless of the structure of the underlying
  network defined by the set of couplings; a general feature that prevents a naive application
  of network theory to quantum systems.

  Our result is obtained by combining the quantum-classical mapping with a random-non-random mapping, the latter being based on a 
  simple topological argument.

A short version of this work was originally submitted to Phys. Rev. Lett. on November 2019.
After completion of the present longer version, I came to know that more recently
Phys. Rev. Lett. have published an interesting work that by simulation shows
the absence of the small-world effect in a photonic quantum network~\cite{Chaves}. 
  
  \begin{acknowledgments}
Grant CNPq 307622/2018-5 is acknowledged.  
We thank T. Macr\`{i} and C. Presilla for a critical reading.

 \end{acknowledgments}

\newpage

\begin{widetext}
\appendix

\section{RNRM}

The RNRM was originally developed for analyzing spin glass models but in the case of ferromagnetic models its derivation
is simpler and it is worth reporting it here where we check its validation for the model with
the Hamiltonian given by Eq. (\ref{Hcl}).
For more details we refer the reader to Refs. \cite{MOI,SW}. We report also a new derivation not based on the replica trick.
Let us consider an Ising model with $N$ spins and generic couplings
$J_b$ along the bonds $b\equiv (i_b,j_b)$ of an undirected graph $G$ whose set of links is denoted by $\Gamma$. The Hamiltonian
and the partition function of this system read as follows:
\begin{eqnarray}
  \label{Happ}
H=-\sum_{b\in\Gamma}J_b\sigma_{i_b}\sigma_{j_b},
\end{eqnarray}
\begin{eqnarray}
  \label{Zapp}
  Z=\prod_b\cosh(\beta J_b)P,
\end{eqnarray}
\begin{eqnarray}
  \label{Papp}
  P= \sum_{\gamma \in \mathcal{G}}\prod_{b\in\gamma} t_b, \quad t_b\equiv \tanh(\beta J_b),
\end{eqnarray}
where in Eqs. (\ref{Zapp})-(\ref{Papp}) we have applied the so called high temperature expansion, $\mathcal{G}$ being the set of (closed)
multi-polygons $\gamma$ in $G$. From Eq. (\ref{Zapp}) we see that in the thermodynamic limit the density free energy $f$ splits as
\begin{eqnarray}
  \label{fapp}
  -\beta f=\lim_{N\to\infty}\frac{1}{N}\sum_{b\in\Gamma}\log\left[\cosh(\beta J_b)\right]+\varphi, 
\end{eqnarray}
where 
\begin{eqnarray}
  \label{phiapp}
  \varphi= \lim_{N\to\infty}\frac{\log(P)}{N}.
\end{eqnarray}
Clearly, a singular behavior of $f$, if any, can be contained only in $\varphi$. Let us consider now that some disorder is present, either
in the graph $G$, in the couplings, or in both,
and let us indicate by $\overline{\cdot}$ the average over this disorder. In other words, the factors
$t_b$ with $b\in \Gamma$, must be seen as a set of independent random variables (not necessarily identically distributed).
In such a case we are interested
in evaluating
\begin{eqnarray}
  \label{phiappd}
  \overline{\varphi}= \lim_{N\to\infty}\frac{\overline{\log(P)}}{N}=\lim_{N\to\infty}\lim_{n\to 0}\frac{\overline{P^n}-1}{Nn}
\end{eqnarray}
The replica trick consists in evaluating $\overline{P^n}$ for $n$ integer and attempt the analytic continuation toward $n\to 0$. From
Eq. (\ref{Papp}) we have
\begin{eqnarray}
  \label{Pappd}
  \overline{P^n}= \sum_{\gamma_1,\ldots,\gamma_n \in \overline{\mathcal{G}}}\prod_{b_1\in\gamma_1,\ldots,b_n\in\gamma_n} \overline{t_{b_1}\cdots t_{b_n}},
\end{eqnarray}
where $\overline{\mathcal{G}}$ is the natural extension of $\mathcal{G}$ that includes disorder, \textit{i.e.}, the set of multi-polygons
made by bonds $b\in\overline{\Gamma}$, where  
\begin{eqnarray}
  \label{Gammad}
  \overline{\Gamma}=\left\{b=(i_b,j_b): \overline{t_b}\neq 0\right\}.
\end{eqnarray}

Let us analyze the first terms. We have
\begin{eqnarray}
  \label{Pappd1}
  \overline{P^1}= \overline{P}=\sum_{\gamma \in \overline{\mathcal{G}}}\prod_{b\in\gamma} \overline{t_{b}},
\end{eqnarray}
\begin{eqnarray}
  \label{Pappd2}
  \overline{P^2}= \sum_{\gamma_1,\gamma_2 \in \overline{\mathcal{G}}}~~\prod_{b_1\in\gamma_1,b_2\in\gamma_2:~b_1\neq b_2} \overline{t_{b_1}}\cdot \overline{t_{b_2}}
  \prod_{b\in\gamma_1\cap\gamma_2} \overline{t^2_{b}},
\end{eqnarray}
\begin{align}
  \label{Pappd3}
  &\overline{P^3}= \sum_{\gamma_1,\gamma_2,\gamma_3 \in \overline{\mathcal{G}}}~~\prod_{b_1\in\gamma_1,b_2\in\gamma_2,b_3\in\gamma_3:~b_1\neq b_2,b_1\neq b_3,b_2\neq b_3}
  \overline{t_{b_1}}\cdot \overline{t_{b_2}}\cdot \overline{t_{b_3}}
  \nonumber\\
  & \prod_{b\in\gamma_1\cap\gamma_2,b'\in\gamma_3:~b\neq b'} \overline{t^2_{b}}\cdot \overline{t_{b'}}
  \prod_{b\in\gamma_1\cap\gamma_3,b'\in\gamma_2:~b\neq b'} \overline{t^2_{b}}\cdot \overline{t_{b'}}
  \prod_{b\in\gamma_2\cap\gamma_3,b'\in\gamma_1:~b\neq b'} \overline{t^2_{b}}\cdot \overline{t_{b'}}
  \prod_{b\in\gamma_1\cap\gamma_2\cap\gamma_3} \overline{t^3_{b}}.
\end{align}
Let us consider a disorder such that $\overline{t_b}\geq 0$.
Depending on the kind of disorder, $\overline{\Gamma}$ can be "small'' or ``large''. For example, if the disorder involves only
the couplings of a lattice model with a fixed set of bonds $\Gamma$, we have $\overline{\Gamma}=\Gamma$ and the evaluation of
$\overline{P^n}$ may remain very hard for $n\neq 1$, unless $G$ is locally tree-like, for which we have
\begin{eqnarray}
  \label{Pappdn}
  \overline{P^n}\simeq \overline{P}^n.
\end{eqnarray}
If instead the disorder is such that at any lattice point there pass a number
$N^\alpha$ of bonds for some $\alpha>0$, a simple combinatorial argument emerges: given two randomly
chosen multi-polygons $\gamma_1$
and $\gamma_2$ in $\overline{\mathcal{G}}$, the probability that they overlap along some bonds becomes negligible in the limit of large $N$.
Notice that, due to the factors $\overline{t^k_{b}}$ in $\overline{P^n}$, with $k\in\{1,\ldots,n\}$, at high temperature
and in the thermodynamic limit, $\overline{P^n}$ is characterized by only short multi-polygons whose overlap, therefore, tends trivially
to zero
(simply because they are mostly disconnected).
As we decrease the temperature, however, multi-polygons of larger and larger length become important in $\overline{P^n}$.
In fact, for a system characterized by a single coupling value and without disorder, i.e., a single value $t_b=\tanh(\beta J)$,
$\varphi$ is a power series in $\tanh(\beta J)$ as follows
\begin{eqnarray}
  \label{phiseries}
  \varphi=\sum_{l=0}^{\infty}c_l \tanh^l(\beta J),
\end{eqnarray}
where the coefficients $c_l$ take into account the number of multi-polygons of length $l$ (see later for the actual connection
between the $c_l$'s and the numbers $C_l$'s of multi-polygons of length $l$) and therefore grow exponentially with $l$.
As a consequence, the series
has a radius of convergence $R$ determined by the inverse of the ratio of growth of $c_l$, $R^{-1}=\lim_{l\to\infty}c_{l+1}/c_l$.
The radius of convergence determines therefore the critical temperature of the system via the universal equation
\begin{eqnarray}
  \label{critic}
  \frac{1}{\tan(\beta_c J)}=\lim_{l\to\infty}\frac{c_{l+1}}{c_l}.
\end{eqnarray}
An analogous formula 
holds true also for the random system.
Equation (\ref{critic}) tells us that, at the critical point, the relevant information concerns only multi-polygons of infinite length
(later on, we comment about how the value of $\overline{\varphi}$ at the critical point characterizes its value in all
the paramagnetic region).
This observation applies in particular to the evaluation of $\overline{P^2}$ via 
the analysis of two randomly
chosen paths of infinite length.
Whereas it is difficult to evaluate the probability that they overlap in general, it is easy to see that
it goes to zero in many cases of interest by just looking at the possible paths that emanate from the same vertex.
This combinatorial argument allows one to effectively neglect all the terms involving overlaps of bonds in Eqs. (\ref{Pappd1})-(\ref{Pappd3})
which leads again to Eq. (\ref{Pappdn}), 
where the approximation becomes exact in the limit $N\to\infty$, regardless of the presence of short loops and regardless of the value of
$\alpha$. Notice that only for $\alpha=1$ one has the small-world property.

More precisely the mechanism goes as follows. To fix the idea let us consider an Ising model with the coupling $J$ built over
the Gilbert random graph, i.e., the graph where  
the adjacency matrix $c_{i,j}$ is a random variable taking values 0 or 1 with
probabilities $1-c/N$ or $c/N$, respectively. In this case we have $\overline{t^k_b}=\overline{t^k}=(c/N) \overline{\tanh^k(\beta J)}$.
As we have explained in Section IV, in the thermodynamic limit, we are free to evaluate $\overline{f}$ and $\overline{\varphi}$
by either performing the average over all the graph realizations, or by considering a single but infinite
graph (in other words the free energy
is self-averaging). We apply the latter if $c<1$ and the former if $c>1$. If $c<1$ we use the known fact that, in the thermodynamic limit,
the Gilbert random graph has zero clustering coefficients. As a consequence,
for $c<1$ in the thermodynamic limit loops are negligible and we have trivially $\overline{\varphi}=\varphi=0$.
Let us now consider the case 
$c>1$ and let us evaluate $\overline{P^2}$.
From Eq. (\ref{Pappd2}) we see that we have a sum over pairs of multi-polygons which can have a zero,
a full, or a partial overlap between each other.
We can split the sum over all the pairs of multi-polygons $\gamma_1$ and $\gamma_2$ of lengths $l_1$ and $l_2$, respectively, as follows
\begin{eqnarray}
  \label{Pappd2b}
  \overline{P^2}= 
  \sum_{l_1,l_2} \left(\mathcal{N}_0(l_1,l_2) \left(\overline{t}\right)^{l_1+l_2}+
  \mathcal{N}_2(l_1,l_2) \left(\overline{t^2}\right)^{(l_1+l_2)/2}+\ldots\right),
\end{eqnarray}
where $\mathcal{N}_0(l_1,l_2)$ is the number of pairs of multi-polygons which have zero overlap,
$\mathcal{N}_2(l_1,l_2)$ is the number of pairs of multi-polygons which have full overlap, and the dots stands for the rest
of the contributions characterized by a partial overlap. 
To make more manifest the first two kinds of contributions we find 
it convenient to rewrite the sum as
\begin{eqnarray}
  \label{Pappd2c}
  \overline{P^2}=
{  \sum_{l=0}^{\infty} \left(\mathcal{N}_0(l)\left(\frac{c}{N}\tanh(\beta J)\right)^l +
  \mathcal{N}_2(l)\left(\frac{c}{N}\tanh^2(\beta J)\right)^{\frac{l}{2}} +\ldots\right),}
\end{eqnarray}
where $\mathcal{N}_0(l)$ is the total number of multi-polygons of length $l$ which can be formed
by any pair of multi-polygons having zero overlap,
$\mathcal{N}_2(l)$ is the total number multi-polygons formed by pairs of multi-polygons each having length $l/2$ (for $l$ even;
clearly, for $l$ odd we have $\mathcal{N}_2(l)=0$), {and we have made explicit use of $\overline{t^k}=(c/N) \overline{\tanh^k(\beta J)}$}. Notice that here $\overline{\Gamma}$
corresponds to the complete (or fully connected)
graph.
Taking into account that in the complete graph of $N$ nodes the number of paths
of length $l$ is $N^l$, 
we have
$\mathcal{N}_0(l)\simeq (l(l+1)/2) {N}^l$ and $\mathcal{N}_2(l)={N}^{l/2}$. In fact, we can form a global path of length $l$
by combining two non overlapping paths
having lengths $l_1$ and $l-l_1$ respectively, and there are $l(l+1)/2$ such combinations, whereas we can overlap a pair of paths
only if they have the same length.
{In conclusion, the ratio between the second and the first kind of contributions goes like $1/(c^{l/2} l^2)$.}
Similar considerations
hold true for the terms with a partial overlap.
Taking into account now that $c>1$,
we see that, in the thermodynamic limit, for $l\to\infty$ Eq. (\ref{Pappd2c}) can be evaluated by dropping all the terms
except those associated with $\mathcal{N}_0(l)$.

By plugging Eq. (\ref{Pappdn}) in Eq. (\ref{phiappd})
and by using Eq. (\ref{Pappd1}) we finally
arrive at
\begin{eqnarray}
  \label{phiappd1}
  \overline{\varphi}=\varphi_I(\{\overline{t_{b}}\}),
\end{eqnarray}
which tells us that, up to an additive constant, the density free energy $f$ of the random system is equal to the density free energy
$f_I$ of a Ising model whose set of non random couplings $\{J^{(I)}_b\}$ satisfies
\begin{eqnarray}
  \label{RNRM1}
  \tanh(\beta J^{(I)}_b)=\overline{\tanh(\beta J_b)}.
\end{eqnarray}
In particular, for the case of the generalized Gilbert random graph with fixed amplitudes related to the Hamiltonian (\ref{Hcl}),
we see that the topological argument holds true for large $N$ so that in this case we have $\alpha=1/2$ (the total number of nodes
is $MN$ with $M\sim N$, see warning (ii)) and Eq. (\ref{RNRM1})
amounts to (here $(i_1,j_1)$ and $(i_2,j_2)$ represent the positions of the two nodes
between which there is the bond $b$, $J^{(I)}_b\equiv J^{(I)}_{(i_1,j_1;i_2,j_2)}$)
\begin{align}
  \label{RNRM2}
  \tanh(\beta J^{(I)}_{(i_1,j_1;i_2,j_2)})=\left\{
  \begin{array}{l}
   \frac{c}{N}\tanh(\beta J_{0x}+\frac{\beta J}{M})+(1-\frac{c}{N})\tanh(\beta J_{0x}), ~i_2=i_1+1,~j_2=j_1,\\
   \tanh(\beta J_{0y}), ~ j_2=j_1+1,~i_2=i_1,\\
   \frac{c}{N}\tanh(\frac{\beta J}{M}), ~ i_2\neq i_1+1,~j_2=j_1,
  \end{array}
  \right.
\end{align}
which for large $M$ leads to the Ising non random Hamiltonian (\ref{Hcl1}).

It is instructive to derive the RNRM in another way which does not make use of 
the replica trick.
For pedagogical reasons, let us first assume that $\mathcal{G}$ is the square lattice.
In $\mathcal{G}$, the allowed multi-polygons $\gamma$ have lengths
$l=0,4,~6~,8,~\ldots$.
Let us decompose
$P$ in Eq. (\ref{Papp}) (i.e., the singular part of the partition function) as
\begin{align}
  \label{Papp1}
  P= 1+ Q_4+Q_6+Q_8+\ldots,
\end{align}
where we have introduced the sums restricted to multi-polygons of fixed length:
\begin{align}
  \label{Qapp}
  Q_l=\sum_{\substack{\gamma \in \mathcal{G}:\\l(\gamma)=l}}\prod_{b\in\gamma} t_b,
\end{align}
$l(\gamma)$ being the length of the multi-polygon $\gamma$.
We have
\begin{align}
  \label{Papp2}
  \log(P)= Q_4+Q_6+Q_8-\frac{1}{2}\left(Q^2_4+Q^2_6+Q^2_8+\ldots\right) 
  -\left(Q_4Q_6+Q_4Q_8+Q_6Q_8+\ldots\right)+\ldots,
\end{align}
which leads to
\begin{align}
  \label{Papp3}
  \log(P)= Q^{\mathrm(c)}_4+Q^{\mathrm(c)}_6+Q^{\mathrm(c)}_8 +\ldots,
\end{align}
where $Q^{\mathrm(c)}_l$ stands for the sum restricted to connected multi-polygons of fixed length $l$,
a connected multi-polygon being a multi-polygon that cannot be obtained by the product of two or more
(disconnected) multi-polygons (in fact, we could call them just ``polygons''). The first few terms are
\begin{align}
  \label{Qapp2a}
  Q^{\mathrm(c)}_4 &=Q_4, \\
  \label{Qapp2b}
  \quad Q^{\mathrm(c)}_6 &=Q_6, \\
  \label{Qapp2c}
 Q^{\mathrm(c)}_8&=Q_8-\frac{1}{2}Q^2_4.\\
 \ldots \nonumber
\end{align}
Notice that, whereas $Q_l$ does not allow for overlaps of links, $Q^{\mathrm(c)}_l$ in general does. 
For example, the case $l=8$ in Eq. (\ref{Qapp2c})
shows that, in $Q^{\mathrm(c)}_8$, the contributions due to the product of two disconnected squares coming from the
second term, cancel out with the first term. 
However, not all the contributions coming from the second term cancel out with some of the first. 
In fact, such extra contributions are all those in which a partial or total overlap between two squares is present.
Note that the most important feature of the $Q^{\mathrm(c)}_l$'s is that they are extensive in the system size $N$,
as it must be according to Eq. (\ref{phiapp}). Indeed, in this framework we understand
the previously mentioned connection between the coefficients $c_l$ introduced in Eq. (\ref{phiseries}), and the total number of
multi-polygons of length $l$, $C_l$: for a system characterized by a single coupling the relation
between the $c_l$'s and the $C_l$'s is the same as the relation between the $Q^{\mathrm(c)}_l$ and the $Q_l$'s.
Let us now suppose that the couplings in Eq. (\ref{Happ}) are independent identically distributed random variables.
From Eqs. (\ref{Papp3}) and (\ref{Qapp2a}-\ref{Qapp2c}) we have
\begin{align}
    \label{Papp4}
  \overline{\log(P)}= \overline{Q^{\mathrm(c)}_4}+\overline{Q^{\mathrm(c)}_6}+\overline{Q^{\mathrm(c)}_8} +\ldots,
\end{align}
where
\begin{align}
  \label{Qapp2aa}
  \overline{Q^{\mathrm(c)}_4} &=\overline{Q_4}, \\
    \label{Qapp2ab}
    \quad \overline{Q^{\mathrm(c)}_6} &=\overline{Q_6}, \\
      \label{Qapp2ac}
 \overline{Q^{\mathrm(c)}_8}&=\overline{Q_8}-\frac{1}{2}\overline{Q^2_4},\\
 \ldots \nonumber
\end{align}
Of course $\overline{Q^2_4~}\neq (\overline{Q_4})^2$, and similarly for higher order terms. However, we recognize that
the approximation $\overline{Q^2_4~}\simeq (\overline{Q_4})^2$
becomes more and more accurate as we consider, instead of the two-dimensional lattice, a hypercube D-dimensional
lattice with larger and larger values of D
(for the same topological argument that we have previously applied in deriving Eq. (\ref{Pappdn})).
We stress again that
both before and after averaging over the disorder, in $Q^2_4$ there are contributions that do not cancel out with some
of those of $Q_8$ and that the former are those in which at least an overlap of two links is present. As we consider larger
and larger values of D, such surviving contributions from $\overline{Q^2_4~}$ become dominated by those pairs of squares
where only a pair of links overlap with each other
(more precisely, by neglecting the contribution with the full overlap with respect to the contributions with
a single overlap, we make an error order 1/D). Now, this does not imply yet that, in evaluating $\overline{Q^{\mathrm(c)}_8}$ for large D,
$\overline{Q^2_4~}$ and $(\overline{Q_4})^2$ can be taken as approximately equal, because of the presence of the above
overlapping link which is associated with a term that does not cancel out with $\overline{Q_8}$.
However, as we have discussed before, in the thermodynamic limit 
the leading contributions that characterize the critical behavior of the system are those
associated with arbitrarily long multi-polygons, i.e., $l\to \infty$, where a single link overlap does not play any role. 
And the
above argument can be repeated for any products of say $k$ terms,
$Q_{l_1}Q_{l_2}\ldots Q_{l_k}$.
This shows that, effectively, when
$D\to\infty$, or, more in general, when $\overline{\mathcal{G}}$ (see definition (\ref{Gammad}))
is such that any vertex has a number of neighbors
which is an increasing function of $N$ (as in our target model), we can take effectively
$\overline{Q_{l_1}Q_{l_2}\ldots Q_{l_k}} \simeq \overline{Q_{l_1}}\cdot \overline{Q_{l_2}}\cdots \overline{Q_{l_k}}$, which, by plugging
in Eqs. (\ref{Qapp2aa}-\ref{Qapp2ac}) implies
\begin{align}
  \label{Qapp3}
  \overline{Q^{\mathrm(c)}_l} \simeq Q^{\mathrm(c)}_l(\{\overline{t_b}\}),
\end{align}
where $Q^{\mathrm(c)}_l(\{\overline{t_b}\})$ stands for the contribution of the connected multi-polygons of length $l$
of a non random system in which the random couplings $\{t_b\}$ are replaced by their averages over the disorder $\{\overline{t_b}\}$
and the approximation becomes exact in the thermodynamic limit. Equation (\ref{Qapp3}) leads immediately to (\ref{phiappd1}).
This alternative derivation,
despite being a little more complicated, shows that actually the RNRM can be proved without invoking any replica trick.

Above, we have proved the RNRM at the critical point, which is enough for the present paper. We point out however that
the same mapping applies to all the paramagnetic region $P$. 
For example, by extending the RNRM to arbitrary disorder, including
spin glass disorder, it is possible to find the generalization of the Nishimori line~\cite{Nishiline} and, by analytic
continuation, to show that $\overline{\varphi}$ takes the same value in all the $P$ region. In particular, for a system with
a fixed coupling built on the Gilbert random graph, this implies that $\overline{\varphi}\equiv 0$ in all the $P$ region~\cite{MOIII}. 
\vspace{4cm}
\end{widetext}


\section*{References}
\begin{thebibliography}{99}%

  \bibitem{Onsager} L. Onsager, 
  Physical Review, Series II, \textbf{65} (3–4): 117–149, (1944).
  
\bibitem{Sondhi} S. L. Sondhi, S. M. Girvin, J. P. Carini, and D. Shahar,
Rev. Mod. Phys. \textbf{69}, 315 (1997).  
  
\bibitem{Sachdev} S. Sachdev, ``Quantum Phase Transitions'', Cambridge University Press (1999).

\bibitem{Vojta} T. Vojta, ``Quantum Phase Transitions'' in: Computational Statistical Physics. Springer, Berlin, Heidelberg (2002).

\bibitem{Essler}  
  F. H. L. Essler, H. Frahm, F. Göhmann, A. Klümper, V. E. Korepin,
  ``The one-dimensional Hubbard model'', Cambridge University Press (2005).
  
\bibitem{Dutta} A. Dutta \textit{et al},
  ``Quantum phase transitions in transverse field spin models: From Statistical Physics to
Quantum Information'', Cambridge University Press (2015). 

\bibitem{QCM} M. Suzuki, Prog. of Theor. Phys. \textbf{56}, 1454 (1976).
  
\bibitem{Parisi} M. Mezard, G. Parisi, M.A. Virasoro,  
\textit{Spin Glass Theory and Beyond} (Singapore: World Scientific) (1987).

\bibitem{Fisher} K.H. Fischer and J.A. Hertz, \textit{Spin Glasses}, Cambridge University Press (1991).
  
\bibitem{Watts} D. J. Watts, S. H. Strogatz, Nature, \textbf{393}, 440 (1998).

\bibitem{Bollobas} B. 
Bollob\'as, ``Random Graphs'' (2nd ed.), Cambridge University Press (2001). 

\bibitem{Zecchina} O. C. Martin, R. Monasson and R. Zecchina,
  ``Statistical mechanics methods and phase transitions in optimization problems'',
Theoretical computer science \textbf{265}, 3-67 (2001).

\bibitem{Barabasi} R. Albert, A.L. Barab\'asi, Rev. Mod. Phys. \textbf{74} 47 (2002).  
  
\bibitem{DM} S.N. Dorogovtsev, J.F.F. Mendes,
\textit{Evolution of Networks} (University Press: Oxford, 2003).

\bibitem{NewBar} M. Newman, A.L. Barab\'asi, D. J. Watts,
\textit{The Structure and Dynamics of Networks} (Princeton Studies in Complexity) (2006). 
 
\bibitem{Guido} G. Caldarelli, \textit{Scale-Free Networks} (Oxford Finance Series) (2007).

\bibitem{Review} S.N. Dorogovtsev, A.V. Goltsev, J.F.F. Mendes,	
Rev. Mod. Phys. \textbf{80}, 1275 (2008).



\bibitem{Biamonte} J. Biamonte, M, Faccin, M. De Domenico, 
  Commun. Phys. \textbf{2}, 53 (2019).

\bibitem{Anderson} C. P. Zhu and S.-J. Xiong, Phys. Rev. B \textbf{62}, 14780 (2000);
  M. Sade, T. Kalisky, S. Havlin, R. Berkovits, Phys. Rev. E, \textbf{72} 066123 (2005);
  L. Jahnke, J. W. Kantelhardt, R. Berkovits, S. Havlin, Phys. Rev. Lett., \textbf{101} 175702 (2008).
  
\bibitem{Entanglement} M. Cuquet, J. Calsamiglia, Phys. Rev. Lett., \textbf{103} 240503 (2009);
  S. Perseguers, M. Lewenstein, A. Ac\i n, J. I. Cirac, J. I., Nature Physics, \textbf{6} 539 (2010).

\bibitem{BianconiSC} G. Bianconi, 
Phys. Rev. E \textbf{85}, 061113 (2012).
  
\bibitem{QuantumInternet} S. Wehner, D. Elkouss, R. Hanson, 
Science \textbf{362}, 303 (2018).
  

\bibitem{Hastings} M. B. Hastings,
  ``Mean-Field and Anomalous Behavior on a Small-World Network,''
  Phys. Rev. Lett. \textbf{91}, 098701 (2003).

\bibitem{Nikoletopoulos} T. Nikoletopoulos \textit{et al}
  J. Phys. A \textbf{37}, 6455 (2004).
  
\bibitem{SW}
M. Ostilli and J. F. F. Mendes,
Phys. Rev. E \textbf{78}, 031102 (2008).

\bibitem{SW2}
A. L. Ferreira, J. F. F. Mendes, and M. Ostilli,
Phys. Rev. E \textbf{82}, 011141 (2010).
  
  
\bibitem{Nishimori} T. Kadowaki and H. Nishimori,
  Phys. Rev. E 58, 5355 (1998).

\bibitem{Farhi} E. Farhi, J. Goldstone, S. Gutmann, and M. Sipser,
  ``Quantum Computation by Adiabatic Evolution'',
  arXiv:quant-ph/0001106 (2000).
  
\bibitem{Farhi2} E. Farhi, J. Goldstone, S. Gutmann, J. Lapan, A. Lundgren,
  and D. Preda,
  Science 292, 472 (2001).

\bibitem{Santoro} G.E. Santoro, E. Tosatti,
  J. Phys. A: Math. Gen. \textbf{39}, R393 (2006).

\bibitem{Morita}  S. Morita and H. Nishimori,
  J. Math. Phys. 49, 125210 (2008).

\bibitem{Das} A. Das and B. K. Chakrabarti,
  Rev. Mod. Phys. 80, 1061 (2008).

\bibitem{Lidar}
  T. Albash and D. A. Lidar,
  ``Adiabatic Quantum Computing'',
  Rev. Mod. Phys. 90, 015002 (2018).  


\bibitem{Battaglia}
D. A. Battaglia, G. E. Santoro, E. Tosatti,
Phys. Rev. E \textbf{71}, 066707 (2005).

\bibitem{Krzakala} T. J\"org, F. Krzakala, J. Kurchan, A. C. Maggs,
Phys. Rev. Lett. \textbf{101}, 147204 (2008).

\bibitem{Krzakala2}  T. J\"og, F. Krzakala, J. Kurchan, A.C. Maggs, J. Pujos,
Europhys. Lett. \textbf{89}, 40004 (2010).
  
\bibitem{Hen}  I. Hen, J. Job, T. Albash, T. F. Rønnow, M. Troyer, and D. A. Lidar,
  Phys. Rev. A \textbf{92}, 042325 (2015).

\bibitem{Denchev} V. S. Denchev, S. Boixo, S. V. Isakov, N. Ding, R. Babbush,
V. Smelyanskiy, J. Martinis, and H. Neven,
Phys. Rev. X \textbf{6}, 031015 (2016).

\bibitem{Mandra} S. Mandr\`a, G. G. Guerreschi, and A. Aspuru-Guzik,
  New J. Phys. \textbf{18}, 073003 (2016).  

\bibitem{Mandra2} S. Mandr\`a and H. G. Katzgraber,
  ``A deceptive step towards quantum speedup detection'' (2017), (arxiv:1711.01368).
  
\bibitem{Mandra3} S. Mandr\` a and H. G. Katzgraber,
  Quantum Sci. Technol. \textbf{2}, 038501 (2017).
  
\bibitem{Katzgraber} H. G. Katzgraber and M. A. Novotny, Phys. Rev. Applied \textbf{10}, 054004 (2018).

\bibitem{QREM} C. Presilla and M. Ostilli, Physica A \textbf{515}, 57 (2019).

\bibitem{MOI}
  M. Ostilli, J. Stat. Mech. P10004 (2006);
  M. Ostilli, J. Stat. Mech. P10005 (2006)
  

\bibitem{Pfeuty} P. Pfeuty,
  Annals of Physics \textbf{90}, 57  (1970).

\bibitem{SuzukiBook}
  S. Suzuki, Jun-ichi Inoue, and B. K. Chakrabarti,
  ``Quantum Ising Phases and Transitions in Transverse Ising Models'',
  (Lecture Notes in Physics), Springer, (2012).


\bibitem{Kramers}
H. A. Kramers and G. H. Wannier
Phys. Rev. \textbf{60}, 252 (1941).  


\bibitem{Kaufman} B. Kaufman,
Physical Review, Series II, \textbf{76} (8): 1232, (1949).

\bibitem{MF} M. Ostilli, EPL, \textbf{97} 50008 (2012).

\bibitem{Nishiline}  H. Nishimori, J. Phys. C: Solid State Phys., \textbf{13} 4071 (1980).

\bibitem{Chaves} S. Brito, A. Canabarro, R. Chaves, and D. Cavalcanti,
  Phys. Rev. Lett., \textbf{124} 210501 (2020).

\bibitem{MOIII}  M. Ostilli, J. Stat. Mech. P09010 (2007).
  
\end{thebibliography}
\end{document}